\documentclass[
    aps,
    prl,
    reprint,
    twocolumn,
    superscriptaddress,
    tightenlines
]{revtex4-2}

\usepackage{amsmath}
\usepackage{amssymb}
\usepackage{mathrsfs}
\usepackage[dvipsnames]{xcolor}
\usepackage{graphicx}
\usepackage{lipsum}
\usepackage[pdfauthor={Filippo Camilloni}]{hyperref}
\usepackage{times}
\usepackage{tikz}
\usepackage{transparent}

\newcommand{\ie}{i.e.,~}
\newcommand{\eg}{e.g.,~}

\hypersetup{
  colorlinks=true, 
  linkcolor=blue, 
  citecolor=magenta, 
}

\definecolor{azureblue}{HTML}{005EFF}
\definecolor{mathGreen}{rgb}{0.14, 0.8, 0.14}
\definecolor{mathBlue}{rgb}{0.4, 0.6, 1}
\definecolor{mathYellow}{rgb}{1, 0.75, 0}
\definecolor{mathRed}{rgb}{0.93, 0.27, 0.27}

\newcommand{\MBHi}{M_{_{\rm BH,0}}} 
\newcommand{\MBH}{M_{_{\rm BH}}} 
\newcommand{\MNS}{M_{_{\rm NS}}} 
\newcommand{\MBNS}{M_{b,{_{\rm NS}}}} 
\newcommand{\JNS}{J_{_{\rm NS}}}
\newcommand{\JBH}{J_{_{\rm BH}}}
\newcommand{\JBHi}{J_{_{\rm BH,0}}}
\newcommand{\RNS}{R_{_{\rm NS}}}
\newcommand{\chiBHi}{\chi_{_{\rm BH,0}}} 
\newcommand{\chiBHf}{\chi_{_{\rm BH}}} 
\newcommand{\chiBH}{{\chi_{_{\rm BH}}}}
\newcommand{\chiNS}{\chi_{_{\rm NS}}} 
\newcommand{\chiKep}{\chi_{_{\rm Kep}}} 

\begin{document}
\title{Spins do not add: why rotating stars do not help spin-up black holes in binaries}
\author{Filippo Camilloni}
\affiliation{Institut f\"ur Theoretische Physik, Goethe Universit\"at,
  Max-von-Laue-Str. 1, 60438 Frankfurt am Main, Germany}

\author{Konrad Topolski}
\affiliation{Hamburg Observatory, Department of Physics, University of
  Hamburg, Gojenbergsweg 112, 21029 Hamburg, Germany}

\author{Keneth Miler}
\affiliation{Institut f\"ur Theoretische Physik, Goethe Universit\"at,
  Max-von-Laue-Str. 1, 60438 Frankfurt am Main, Germany}

\author{Luciano Rezzolla}
\affiliation{Institut f\"ur Theoretische Physik, Goethe Universit\"at,
  Max-von-Laue-Str. 1, 60438 Frankfurt am Main, Germany}
\affiliation{Frankfurt Institute for Advanced Studies,
  Ruth-Moufang-Str. 1, 60438 Frankfurt am Main, Germany}
\affiliation{School of Mathematics, Trinity College, Dublin 2, Ireland}

\begin{abstract}
  It is well-known that in the merger of a black-hole binary with
  spins aligned with the orbital angular momentum, the spin of the
  final black hole always increases with the spin of the
  components. We show that this very intuitive behavior is violated in
  black hole--neutron star binaries, where the spin of the final black
  hole actually decreases as the spin of the neutron star increases.
  Using a large set of binaries with different spins and mass ratios,
  we identify the mechanism behind this surprising result in the
  nonlinear dynamics of the tidal disruption, which drives a
  substantial loss of angular momentum via dynamical ejecta, in
  some cases comparable to that radiated via gravitational waves. Our
  results highlight that the spin evolution of stellar-mass black
  holes involved in binary mergers is different from what normally
  assumed, far more complex, and interesting.
\end{abstract}

\maketitle

\smallskip\noindent\emph{Introduction}---The coalescence of binary
systems of neutron stars (BNS) and of black hole--neutron stars (BHNS)
constitutes some of the most promising astrophysical laboratories for
fundamental physics~\cite{Damour2007, Baiotti2016, Metzger2017}, with the
potential of providing simultaneous constraints on the equation of state
(EOS) of nuclear matter~\cite{Vallisneri00, Haster2020, Abbott2018b},
gravity in highly dynamical and non-linear regimes~\cite{Bauswein2019},
magnetically-driven phenomena~\cite{Anderson2008, Kiuchi2010, Combi2023,
  Musolino2024b}, the nucleosynthesis of the
heavy-elements~\cite{Tanaka2017, Valenti2017, Wanajo2022}, and even
cosmological observables~\cite{Messenger:2011, Messenger2013,
  LIGOScientific:2017adf, Feeney2020}. Extracting this wealth of
information from observations requires accurate theoretical predictions
that can only be obtained with sophisticated general-relativistic
magnetohydrodynamics (GRMHD) simulations coupling accurate spacetime
evolution to magnetized fluid dynamics~\cite{Janka99a, Loeffler06a,
  Etienne08, Etienne2007b, ShibataTaniguchi2008, Giacomazzo:2010,
  Chawla:2010sw, Foucart2013a, Hayashi2021a, Topolski2024b,
  Topolski2024c}.

Gravitational-wave (GW) observations carried out by the
LIGO-Virgo-Kagra collaboration have detected signals from BHNS
binaries (GW200105 and GW200115~\cite{LIGOScientific:2021qlt}) with a
high degree of confidence, along with other potential
candidates~\cite{LIGOScientific:2020ibl, LIGOScientific2025,
  LIGOScientific2026}, signalling the need for waveform models that
are fast and accurate over a broad range of binary parameters.
Together with their ability to provide information on the EOS~(see, \eg
\cite{Pannarale2011}) BHNS binaries are promising sources for
multimessenger astrophysics as progenitors of tidal-disruption events
(TDEs), transient phenomena that occur when compact stars enter tidal
fields sufficiently strong to overcome their self-gravity. Upon
disruption, most of the NS is accreted by the BH, efficiently increasing
its mass and spin on short timescales ($\sim 1\, \text{ms}$), whereas a
fraction of the initial baryon mass forms long-lived ($\gg 10\,
\text{ms}$) accretion disks and dynamical ejecta. Under these conditions,
BHNS mergers can become electromagnetically bright, owing to
accretion-powered relativistic jets, as well as mildly relativistic
dynamical ejecta and secular winds, leading to kilonova
emissions~\cite{Ruiz2018, Bulla2020, Kyutoku2021a, Most2021a,
  Hayashi2022, Gottlieb2023a, Hayashi2023, Topolski2024d, Markin2026}.

BHNS phenomenology is conveniently classified through the ratio between
the tidal-disruption radius and the radius of the innermost stable
circular orbit, $\xi: = r_{_{\rm TD}}/r_{_{\rm
    ISCO}}$~\cite{Pannarale2010, Kyutoku2011c, Zappa2019}. Type-I ($\xi
\gg 1$) and Type-III ($\xi \approx 1$) BHNSs exhibit strong and moderate
TDEs, whereas Type-II binaries ($\xi < 1$), correspond to plunging
scenarios not substantially different from binary black-holes
(BBHs)~\cite{Foucart2013b}. Both numerical simulations and
semi-analytical considerations indicate that high BH spins, low stellar
compactness, and sufficiently high mass-ratios, favor TDEs over plunge
scenarios.

A natural and almost inevitable question when considering binary mergers
of compact objects concerns the properties of the remnant. For BBHs, this
question regards the mass, spin, and recoil of the final BH and has been
addressed in a large number of works (see, \eg \cite{Rezzolla-etal-2007,
  Rezzolla-etal-2007b, Rezzolla-etal-2007c, DamourNagar:07a,
  Buonanno:07b, Kesden:2008,Lousto:2009mf, Barausse:2009uz, Healy2014,
  Hofmann2016, JimenezForteza2016}). In particular, it is widely accepted
that in the case of BHs whose spins are aligned with the orbital angular
momentum, the spin of the remnant BH \textit{increases} with the spin of the
components. With a few exceptions~\cite{Pannarale2013b, Zappa2019}, an
equivalent knowledge in the case of BHNS binaries is still lacking,
especially for rapidly rotating NSs~\cite{East2015, Ruiz2020b}.

In this \textit{Letter} we fill this gap with a large set of GRMHD
simulations of BHNS binaries and demonstrate that the natural behavior
shown by BBHs is violated in BHNS binaries with comparable masses, where
the spin of the final BH \textit{decreases} as the spin of the NS
increases, so that the largest final spin is reached for nonspinning
NSs. Our findings demonstrate that the NS spin is a non-additive
contribution to the remnant BH spin and has nontrivial effects on the
potential electromagnetic counterparts, making a putative jet dimmer due
to lower remnant spin~\cite{Blandford1977, Meszaros00, Tchekhovskoy2010,
  Tchekhovskoy2011, Camilloni2022, Meringolo2025a}, and the kilonova
brighter due to larger ejecta mass~\cite{Kawaguchi2016, Dietrich2016,
  Metzger2017, Abbott2017c}.

\smallskip\noindent\emph{Parameter space}---Since nondisruptive BHNS
binaries behave very similarly to BBHs, for which analytic expressions
already provide a good approximation, the interesting part of our space
of parameters necessarily comprises only those BHNSs that undergo
TDEs. The general space of parameters comprises the mass-ratio $q :=
\MNS/\MBHi$, the initial BH spin, $\chiBHi := \JBHi/\MBHi^{2}$, the NS
spin, $\chiNS := \JNS/\MNS^{2}$, and the stellar compactness $\mathcal{C}
:= \MNS /\RNS$. To maximise the spin-gain of the BH remnant, we consider
BHNS binaries with $q\lesssim 1$ and irrotational BHs, \ie
$\chiBHi=0$. While this choice slightly disfavors tidal disruptions, it
highlights the influence of the NS spin~\cite{East2015, Ruiz2020b} over
the widely investigated effect of BH spins~\cite{Etienne:2008re,
  Foucart2010, Foucart2011, East2012b0, Foucart2012, Foucart2013a,
  Lovelace2013, Lackey2013, Kyutoku2015, Kawaguchi2015, Ruiz2018,
  Chaurasia2021, Chen2024, Gonzalez2025, Matur2026}.
Demanding the NS to be tidally disrupted by nonspinning BHs of comparable
mass further constrains the parameter space, since $\xi \approx
\mathcal{C}^{-1} q^{-2/3}/6 \gtrsim 1$~\cite{Zappa2019} sets an upper
bound on the compactness, $\mathcal{C} \lesssim 1/6$, requiring stiff
EOSs. As a useful reference, we here employ the DD2
EOS~\cite{Typel:2009sy, Hempel:2009mc}, so that the maximum NS mass and
spin compatible with the compactness limit given above are $\MNS\lesssim
1.5\, M_{\odot}$ and $\chiNS\lesssim 0.7$, respectively.
We fix $\MNS=1.35\,M_\odot$ and endow the star with a poloidal magnetic
field with maximum strength $B_{0}\approx 10^{16}\, {\rm G}$. While this
is sufficiently strong to explore a magnetically-dominated post-merger
emission~\cite{Hayashi2021a}, it does not affect the orbital
dynamics~\cite{Giacomazzo:2009mp}, nor bias our diagnostics for the
remnant BH, disk and dynamical ejecta~\cite{Etienne2012}. Overall, our
set of initial data comprises a total of $16$ BHNS binaries representing
the largest simulation campaign of low-mass BHNS mergers in
quasi-circular orbits that combine a realistic EOS, magnetic fields, and
very rapid stellar rotation (see Table~\ref{table:1}).

\smallskip\noindent\emph{Tidal-disruption phenomenology}---The evolution
is performed using the numerical infrastructure described in
Refs.~\cite{Topolski2024b, Topolski2024c, Topolski2024d}, where the
reader can find details. Given the relatively short timescales
investigated here and the rather low temperature achieved in the
disruption, we do not include neutrino emission and irradiation
(see~\cite{Topolski2024d} for a discussion).

\begin{figure}
  \centering
  \includegraphics[width=\columnwidth]{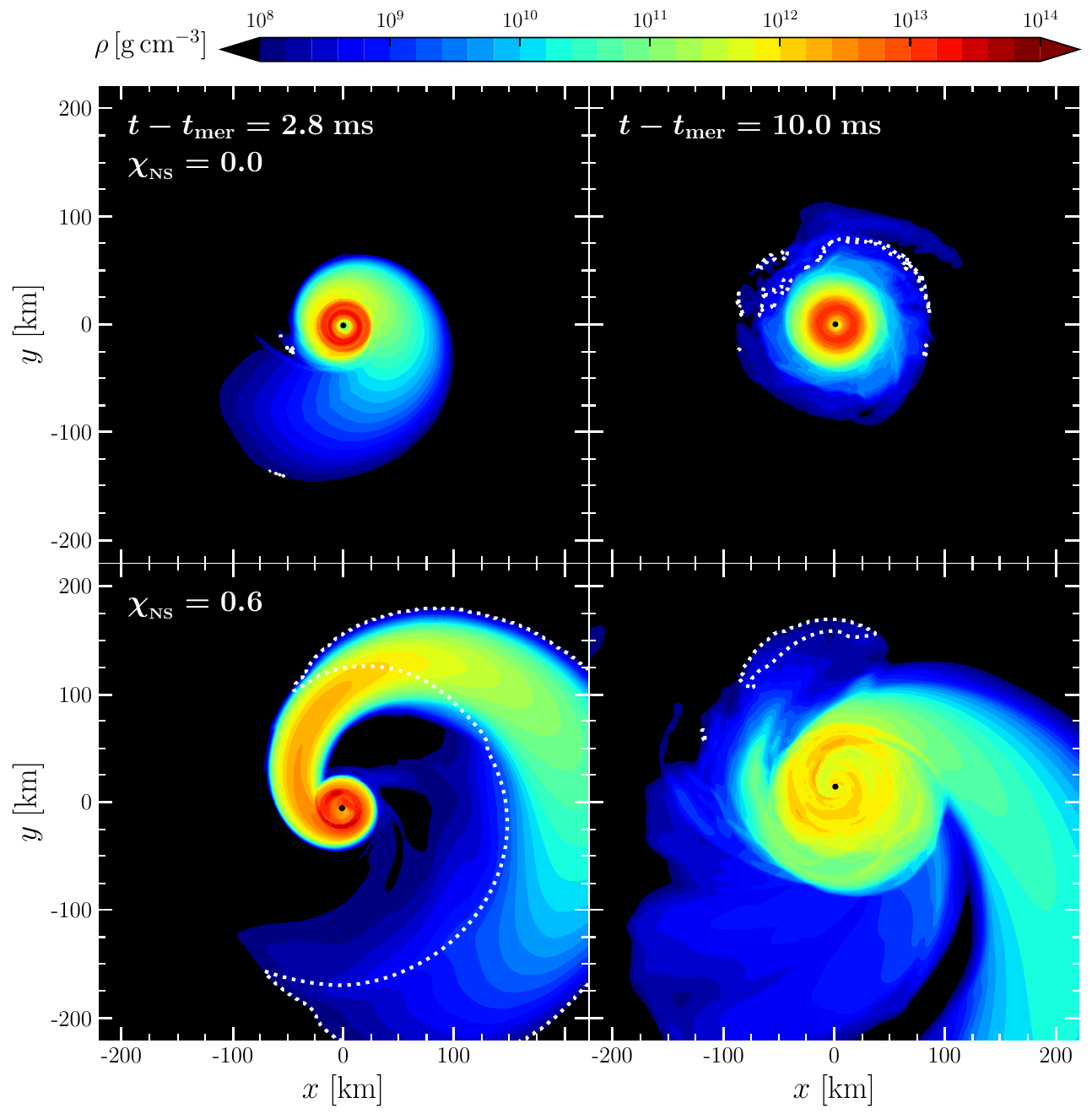}
  \caption{Snapshots of the rest-mass density $\rho$ for equal-mass
    binaries and $\chiNS=0.0$ (top row) and $\chiNS=0.6$ (bottom row).
    The left column refers to the early post-merger phase, \ie $t -
    t_{\rm mer} = 2.8\, {\rm ms}$, while the right one to the late
    post-merger phase, $t-t_{\rm mer} = 10 \, {\rm ms}$ (the data is
    rotated to align the spiral arms). The dashed white contour separates
    bound and unbound ejecta.}
  \label{fig:evolution}
\end{figure}

\begin{figure}
  \centering \includegraphics[width=0.95\columnwidth]{
    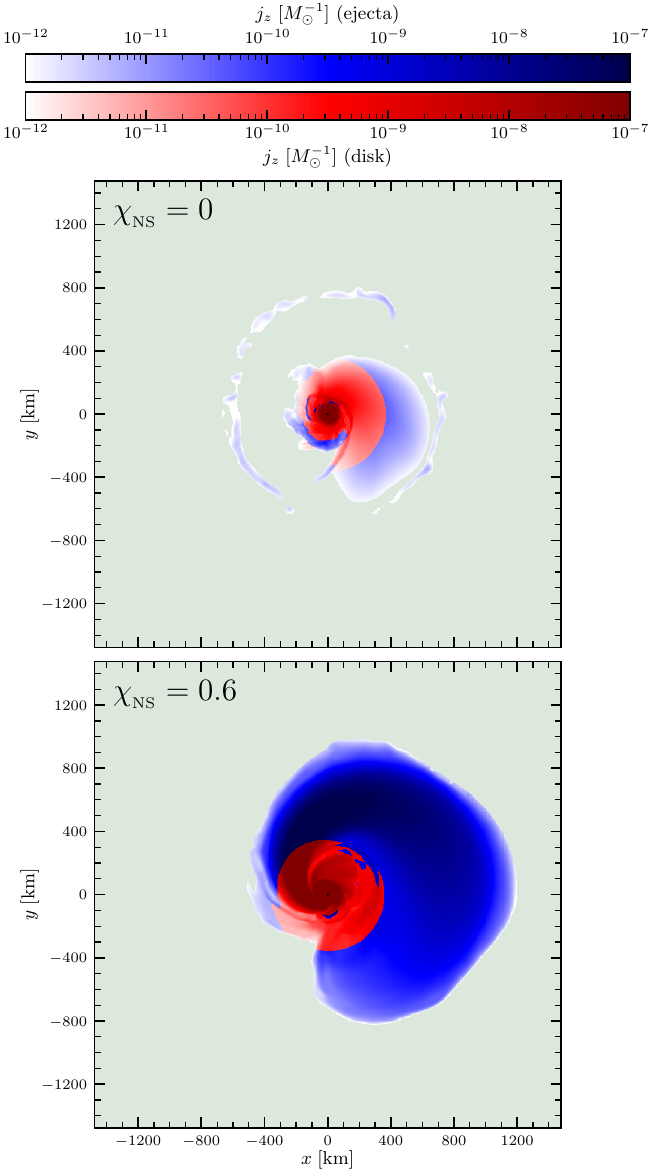}
  \caption{Orbital-plane ($z=0$) projections of the Komar
    angular-momentum density $j_{z}$. A red (blue) colormap is used to
    refer to matter that is bound, $h u_{t} < -1$ (unbound, $h u_{t} >
    -1$) The two snapshots refer to equal-mass binaries at $t-t_{\rm
      mer}=10\, {\rm ms}$ where the NS is either non-spinning (top panel)
    or rapidly spinning (bottom panel). }
  \label{fig:Komar}
\end{figure}

Figure~\ref{fig:evolution} provides snapshots of the rest-mass density
$\rho$ for the binaries with $q=1,\, \chiNS=0.0$ (top row) and $q=1,\,
\chiNS=0.6$ (bottom row), revealing striking differences between BHNS
binaries with irrotational and rapidly rotating NSs. The left column
refers to the early post-merger phase when the NS has just been
disrupted, \ie $t - t_{\rm mer} =2.8\, {\rm ms}$, where $t_{\rm mer}$ is
the merger time and is defined as the maximum of the gravitational-wave
strain. The right column, on the other hand, depicts the late post-merger
phase, $t-t_{\rm mer} = 10 \, {\rm ms}$, when the spacetime attains
stationarity and the disk is completely formed. Overall,
Fig.~\ref{fig:evolution} illustrates how the NS spin significantly
affects the amount and distribution of rest-mass outside the apparent
horizon, highlighting that the NS with an intrinsically high spin will
shed a considerable portion of its mass into a spiral arm that extends up
to the edge of domain. Of course, not all of the mass shed is actually
unbound and will actually contribute to the formation of an accretion
disk (see Fig.~\ref{fig:Komar} and discussion below).

The dashed white line marks the relativistic Bernoulli-criterion boundary
$h u_{t} = -1$~\cite{Rezzolla_book:2013, Bovard2016}, separating the
matter that is bound ($h u_{t}<-1$) and will produce an accretion disk,
from that reaching spatial infinity ($h u_{t} < -1$).
This helps appreciate that, unlike irrotational NSs, rapidly rotating NSs
at tidal disruption can eject a considerable fraction of their mass to
large distances while still contributing to the construction of massive
accretion disks. This is because for $\chiNS>0$, stellar disruption
occurs at larger radii, due to the combined effect of a lower stellar
compactness and a repulsive spin-orbit interaction~\cite{Blanchet2014,
  Campanelli:2006vp}. The material in this case spreads over larger
volumes, leading to the formation of magnetized accretion disks with
$M_{\rm disk}> 0.1 \, M_{\odot}$, sufficient for the post-merger remnant
to power relativistic jets and short gamma-ray bursts on timescales of
$\mathcal{O}(100)$ ms~\cite{Chawla:2010sw, Rezzolla:2011,
  Ruiz2020a}. While Fig.~\ref{fig:evolution} specializes to the most
diverse binaries considered, all simulations exhibit similar qualitative
features when compared across different NS spin at fixed
mass-ratios. Furthermore, for $\chiNS=0.6$, we find that the mass in the
bound disk is $M_{{\rm disk}}\simeq 0.14 \, M_{\odot}$ for $q=1$ (and the
spin-orbit coupling is more effective), whereas $M _{{\rm disk}}\simeq
0.11 \, M_{\odot}$ when $q=0.4$. Crucially, $\chiNS>0$ is necessary for
the dynamical ejecta masses to be well above the $10^{-3}M _{\odot}$
threshold, and thus capable of producing detectable transient kilonovae
signatures~\cite{Metzger2017}.

As discussed below, this markedly different behavior has a direct impact
on the remnant BH. We should also note that while our results confirm
that low-mass BHNS binaries with $\chiNS=0$ produce small amounts of
remnant baryon-mass~\cite{Foucart2019}, a systematic bias exists in the
analytical formulas estimating these masses since they do not account for
the spin of the NS~\cite{Foucart2018b}. Fortunately, quasi-universal
relations on the oblateness of rotating NSs~\cite{Musolino2023b} can be
helpful to introduce simple analytic corrections that improve the
analytic fits currently used to trigger electromagnetic
follow-up~\cite{Foucart2012,Pannarale2014,Foucart2018b} (see EM).

\begin{figure}[t!]
  \centering
  \includegraphics[width=1\columnwidth]{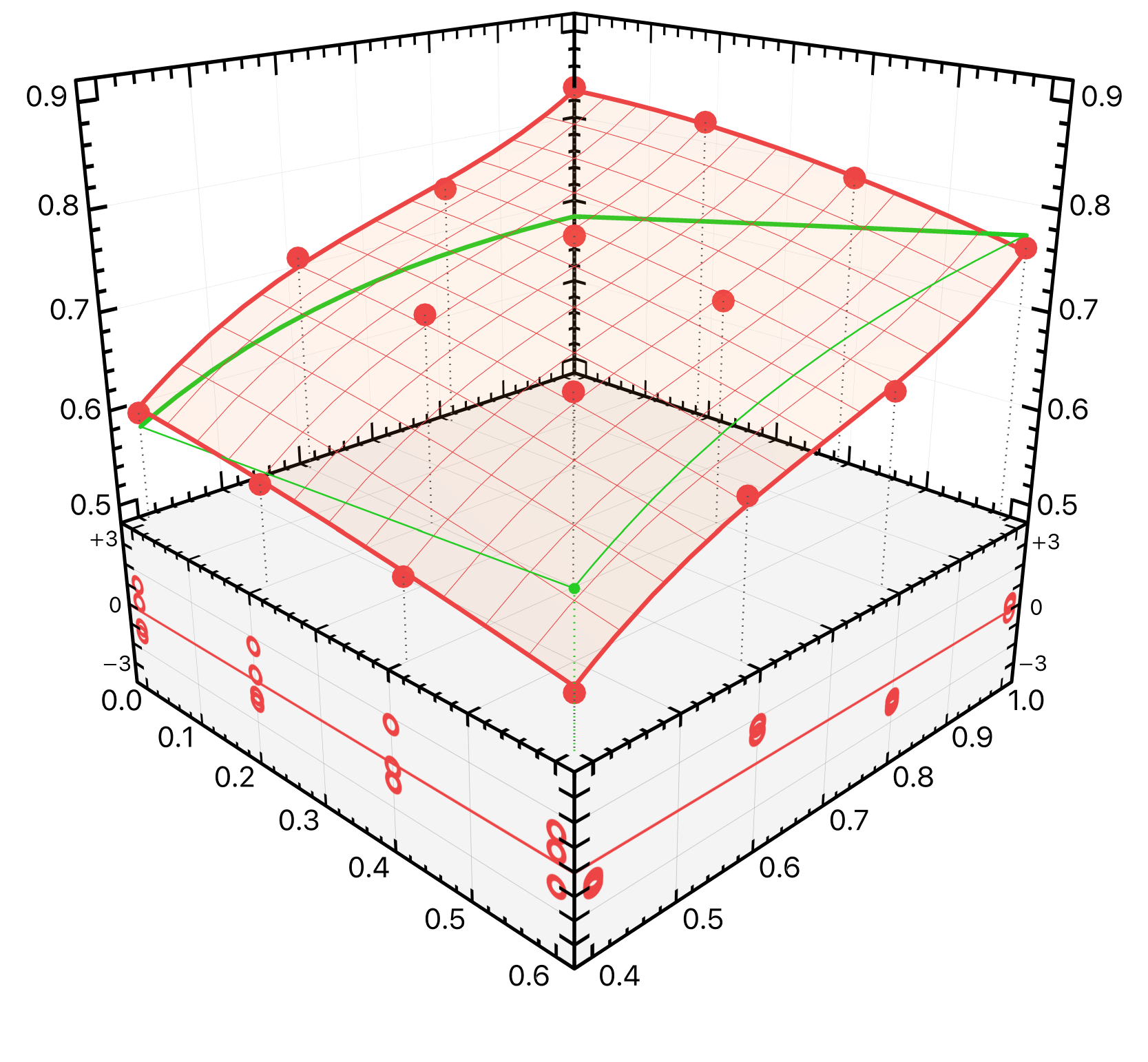}
  \begin{picture}(0, 0)
    \put(-125, 203){\rotatebox{7}{\large{$\chiBH$}}} \put(110, 203){\rotatebox{-7}{\large{$\chiBH$}}}
    \put(-75, 55){\rotatebox{-35}{\large{$\chiNS$}}} \put(60, 50){\rotatebox{45}{\large{$q$}}}
    \put(82, 193){\transparent{1.0}\color{mathGreen} \rotatebox{0}{BBH}}
    \put(108, 100){\scriptsize\rotatebox{86}{$\delta\tiny{[\%]}$}}
    \put(-114,100){\scriptsize\rotatebox{94}{$\delta\tiny{[\%]}$}}
  \end{picture}
  \vspace{-1cm}
  \caption{Dependence of the BH spin $\chiBH$ on the mass ratio $q$ and
    the NS spin $\chiNS$. The red filled circles represent the results of
    our simulations (Table~\ref{table:1}), whereas the surface is the
    polynomial fit. The thick and thin green lines report the
    corresponding behavior for BBH mergers~\cite{
      Rezzolla-etal-2007c}. The lower shows the quality of the fit by
    reporting the relative deviation between the data and the fit
    ($\delta\lesssim 2\%$). }
  \label{fig:fitchi}
\end{figure}

\smallskip\noindent\emph{Black-hole remnant properties}---When the NS is
tidally disrupted, the BH rapidly accretes a large fraction of its mass
and spin, significantly impacting the properties of the BH remnant, which
is the main focus of our work. Both the final BH mass $\MBH$ and spin
$\JBH$, are computed via quasi-local measurements~\cite{Ashtekar03a,
  Dreyer02a} at the reference time, $t =t_{\rm mer}+ 10 \, {\rm
  ms}$~\cite{Foucart2019} (see Table~\ref{table:1}). From these
quantities, we derive the dimensionless BH spin $\chiBH := \JBH/\MBH^2$,
which we report in Fig.~\ref{fig:fitchi} with filled red circles as a
function of the mass-ratio $q$ and of the NS spin $\chiNS$ (we recall
that $\chiBHi=0$).

The behaviour of $\chiBH$ shown in Fig.~\ref{fig:fitchi} is both expected
and unexpected. On rather general terms, we expect that $\chiBH = \chiBH
(q, \chiNS, {\rm EOS})$ and that, once the EOS is fixed, $(\partial
\chiBH/\partial q)_{\chi_{_{\rm NS}}} > 0$ and $(\partial \chiBH/\partial
\chiNS)_q >0$, \ie that the final spin increases both with the mass-ratio
and with the total spin of the binary. This expectation comes from BBHs,
where this behaviour has been first measured almost twenty years
ago~\cite{Rezzolla-etal-2007, Rezzolla-etal-2007b, Rezzolla-etal-2007c}
(see green lines and circle in Fig.~\ref{fig:fitchi}). In the case of
BHNS binaries, however, it is clear from Fig.~\ref{fig:fitchi} that this
no longer true and that while the final spin increases with mass ratio,
it decreases with spin, \ie $(\partial \chiBH/\partial q)_{\chi_{_{\rm
      NS}}} > 0$ and $(\partial \chiBH/\partial \chiNS)_q <0$.

The smooth and regular behavior of the data and the additional constraint
that no spin-up can take place in the extreme-mass-ratio limit, \ie
$\chiBH=0$ for $q \to 0$, allows us to represent the simulation data with
a double polynomial expansion truncated at third order in both $q$ and
$\chiNS$, namely $\chi_{_{\rm BH}}\!=\!b_{1} q +b_{2} q^{2}+b_{3}
q^{3}+c_{1} q \chiNS + c_{2} q^{2}\chiNS+c_{3} q \chiNS^{2}$, where
$b_{1}=2.486\pm 0.049$, $b_{2}=-2.983\pm 0.129$, $b_{3}=1.333\pm 0.085$,
$c_{1}=-0.266\pm 0.073$, $c_{2}=-0.173\pm 0.073$, and $c_{3}=0.242\pm
0.069$. Considering that each simulation data-point has a relative error
measured to be $\Delta\chiBH/\chiBH \simeq 1\%$, the fit represents the
data with very precisely, with deviations $\delta\lesssim 2\%$.

\begin{figure}
  \centering
  \includegraphics[width=\columnwidth]{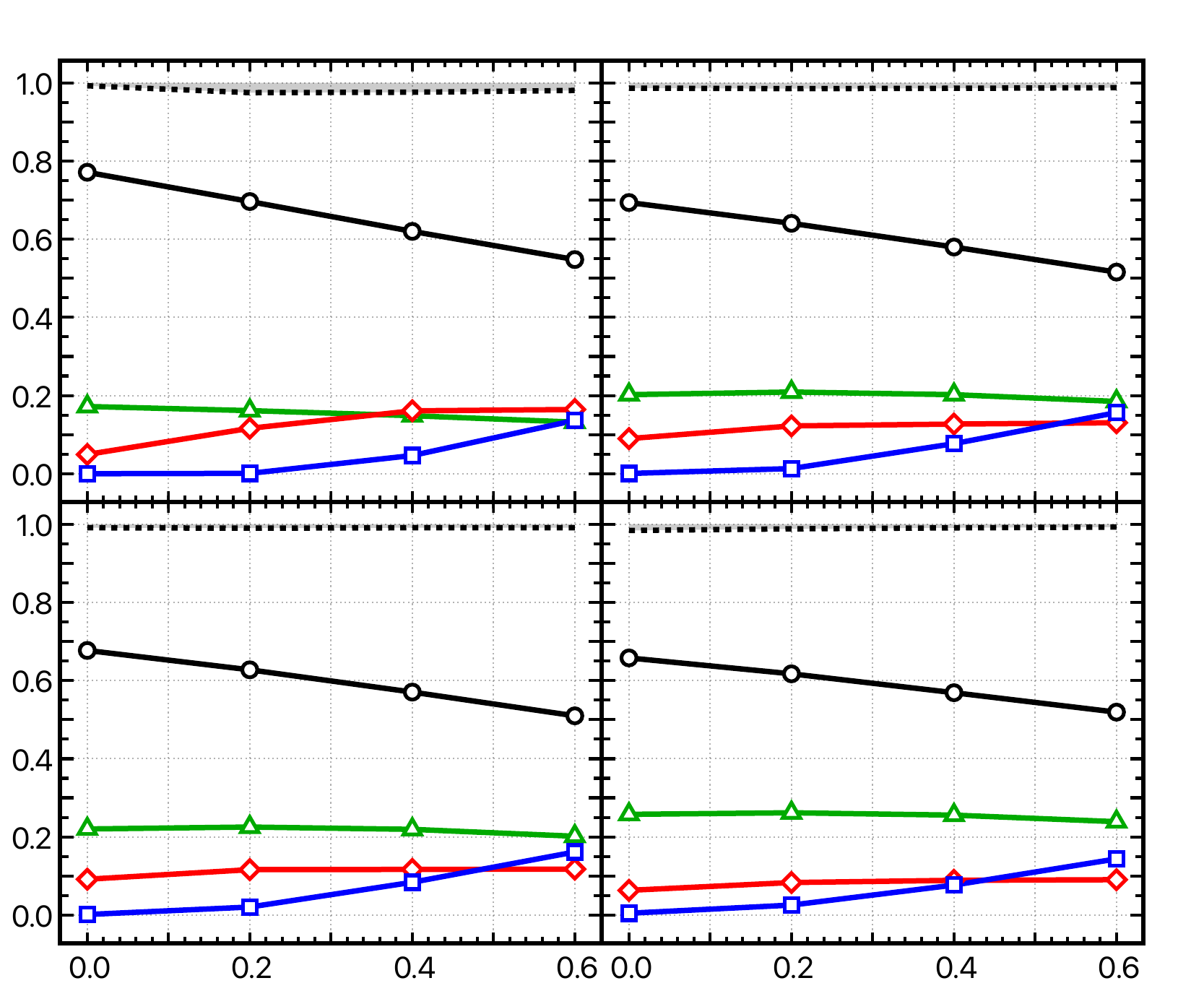}
  \begin{picture}(0, 0)
    \put(-100, 183){\footnotesize {$\tilde{J}_{_{\rm BH}}$}} \put(-100,
    137){\footnotesize \textcolor{Green}{$\tilde{J}_{_{\rm GW}}$}}
    \put(-25, 137){\footnotesize \textcolor{red}{$\tilde{J}_{\rm disk}$}}
    \put(-18, 119){\footnotesize \textcolor{blue}{$\tilde{J}_{\rm ej}$}}
    \put(-60, 10){$\chiNS$} \put(50, 10){$\chiNS$} \put(-36, 183){ {$q=1.0$}}
    \put(75, 183){ {$q=0.8$}} \put(-36, 93){ {$q=0.6$}} \put(75, 93){ {$q=0.4$}}
  \end{picture}
  \caption{Decomposition in the various components of the total angular
    momentum of the system at $10\, \rm{ms}$ after the merger,
    $\tilde{J}_{n}:= J_{n}/J_{0}$, where $n=\{{\rm BH,\, GW,\, disk,\,
      ej}\}$ and $\tilde{J}_{0} = \sum_n \tilde{J}_{n}$ (black dashed
    line). The sum agrees with the initial angular momentum to within
    $0.8$-$2.5\%$ (gray-shades above the dashed curve). }
  \label{fig:balance}
\end{figure}

It is important to remark that, for irrotational NSs, BHNS binaries are
more efficient than BBHs at spinning up the final remnant. There are two
justification behind this result. First, being more compact, BHs are much
more efficient than NSs in radiating the orbital angular momentum, so the
net orbital angular momentum imparted to the remnant is
smaller~\cite{Barausse:2009uz}. Second, the tidal disruption of the NSs
at large distances from the horizon provides the accreting matter with
(orbital) angular momentum that is transferred to the BH, leading to a
more efficient spin-up.

Another remarkable behaviour of the remnant dimensionless spin reported
in Fig.~\ref{fig:fitchi} is summarised in our title, namely that, for any
mass ratio, the \emph{NS spin does not help spin-up BH remnants}. The
markedly different behaviour from the pure-vacuum case and the
non-additivity of spins in BHNS binaries clearly hints at the crucial
role played by matter in TDEs and is directly reflected by the disruption
phenomenology outlined above. More specifically, since aligned NS spins
favor the launching of strong dynamical ejecta and the formation of
massive accretion disks, the BH spin-up is quenched by a good fraction of
the angular momentum remaining in the plasma outside the horizon. This
conclusion can be verified by decomposing the total angular momentum,
which is initially $J_{0} = J_{_{\rm orb}}+\MNS^{2}\chiNS$ into the
different components appearing after the TDE, namely, the BH spin
$J_{_{\rm BH}}$, the angular momentum in the GW field $J_{_{\rm GW}}$,
that in the accretion disk $J_{_{\rm disk}}$, and the one in the ejected
matter $J_{\rm ej}$.

Figure~\ref{fig:balance} tracks with different colors the individual
normalised contributions, $\tilde{J}_{n} := J_{n}/J_0$, where $n=\{{\rm
  BH,\, GW,\, disk,\, ej}\}$, as a function of the NS spin and for
different mass ratios and validates angular momentum conservation, \ie
$J_{0}=J_{_{\rm BH}}+J_{_{\rm GW}}+J_{\rm disk}+J_{\rm ej}$, to a
relative precision of $\lesssim 2.5\%$, consistent with similar
measurements for BNS binaries~\cite{Kiuchi2026a}. As expected if the
novel final-spin behavior is to be found in the dynamics of the TDE
event, any increase in the binary asymmetry, \ie $q \leq 1$, comes with
an increase in the angular momentum lost in GWs corresponding to the
greater binary total mass. At the same time, GWs are not always the
dominant channel of angular-momentum transfer for rapidly spinning
NS. For example, for $q=1$ and $\chiNS=0.6$, the disk retains the
majority of the angular momentum deposited outside the horizon,
$\tilde{J}_{\rm disk} \approx 16.4\%$, whereas the ejecta and the GWs
transport the same amount of angular momentum away from the system,
$\tilde{J}_{\rm ej} \simeq 13.6\%$ and $\tilde{J}_{_{\rm GW}}\simeq
13.2\%$. Furthermore, when $q<1$, the angular-momentum fraction in the
disk does not significantly depend on $\chiNS$, whereas the fraction
transported by the ejecta sharply increases from $\tilde{J}_{\rm
  ej}\lesssim 2\%$ when $\chiNS=0.2$, to $\tilde{J}_{\rm ej} \approx
15\%$ when $\chiNS=0.6$.

Figure~\ref{fig:Komar}, reports the orbital-plane ($z=0$) projection of
the Komar angular-momentum density $j_{z} := n_{\mu}
(\partial_{\phi})^{\nu} T^{\mu}_{~\nu}$ (here, $T^{\mu}_{~\nu}$ is the
energy-momentum tensor and $n_\mu$ the normal-observer co-vector), where
care has been paid in distinguishing the integration for the matter that
is bound and will therefore end-up building the disk ($h u_{t} < -1$; red
colormap) from the one that is unbound and will be instead lost to
spatial infinity ($h u_{t}> -1$; blue colormap). The two snapshots refer
to two equal-mass binaries at $t-t_{\rm mer}=10\, {\rm ms}$ and highlight
that for non-spinning NS (top panel) $\tilde{J}_{\rm disk} \gtrsim
\tilde{J}_{\rm ej} \ll 1$, \ie the angular momentum in the matter (either
bound or unbound) is small, while in the case of the most rapidly
spinning NS (bottom panel) $\tilde{J}_{\rm disk} \gtrsim \tilde{J}_{\rm
  ej} \simeq 0.15$, \ie the matter angular momentum is substantial and
spreads over larger volumes.

Finally, we comment on the future spin evolution of the BHNSs considered
here. Of course the electromagnetic emission from BHNS mergers depends
crucially on $\chiBH$ over the timescales $t-t_{\rm mer}=
\mathcal{O}(10)\, {\rm ms}$ explored with our simulations. Yet, it is
reasonable to ask what would happen to the BH-spin over the much longer
viscous-accretion timescale, \ie for $t-t_{\rm mer}=
\mathcal{O}(1\!-\!10)\, {\rm s}$. While it is not yet possible to carry
out full numerical-relativity simulations over these timescales, it is
possible to show that even when allowing for additional disk mass and
angular momentum to be accreted onto the BH assuming realistic losses to
infinity~\cite{Hayashi2023, Gottlieb2023}, it remains true that spins do
not add in BHNS mergers (see EM).

\smallskip\noindent\emph{Conclusions}---We have presented the first
campaign of numerical-relativity simulations systematically
investigating the BH spin-up in low-mass BHNS binaries experiencing a
TDE. The previously unexplored parameter space considered here is
characterized by low mass-ratios, $\MNS/\MBH\geq 0.4$, nonspinning
BHs, $\chiBH=0$, and highly spinning NSs, $\chiNS\leq0.6$, with spins
aligned with the orbital angular momentum. This allows us to consider
the disruption dynamics of a NS described by a realistic
temperature-dependent EOS and with the highest spin considered in the
context of BHNS mergers~\footnote{Values of $\chiNS\approx 0.67$ have
been recently considered in BNS mergers~\cite{Karakas2024}, whereas
$\chiNS\approx 0.756$ has been reached in BHNS mergers with
polytropes~\cite{East2015}.}.

Among the various findings of this campaign, the most interesting is
that, in BHNS binaries, the spin of the NS does not ``add'' to the
orbital angular momentum, as one might naively expect from BBH intuition,
where fast-spinning progenitors produce faster spinning
remnants. Instead, for all mass ratios we find $(\partial \chiBH/\partial
\chiNS)_q <0$, with the fastest-spinning BH remnants descending from
irrotational NS progenitors and reaching final spins that are even larger
than those of equivalent BBH systems.

The physical origin of this counterintuitive result is to be found in the
analysis of the angular momentum budget, which reveals that the
non-negligible spin-orbit interaction, together with the proximity of the
NS to its mass-shedding limit, leads to large tidal-disruption radii. As
a result, a significant amount of orbital angular momentum is deposited
in matter ejected along fast and unbound trajectories. In turn, this
ejection and angular-momentum loss reduces the orbital angular momentum
available for accretion onto the BH, even on much longer timescales than
those modelled here.

It is worth stressing that this is not a minor quantitative correction to
an intuitive behavior, but a qualitative reversal that reflects the
intricate interplay between matter dynamics and spin-orbit coupling that
is unique to low-mass BHNS systems undergoing TDEs. The implications of
these findings are numerous and impact both population-synthesis studies
-- which have so far treated the BH-spin evolution in BHNS mergers with
the phenomenology of BBHs~\cite{Belczynski08, Chattopadhyay2022,
  Zappa2019} -- and the modeling of multimessenger observations -- where
the spin of the remnant BHs in BHNS mergers has been incorrectly estimated.

Besides providing the first evidence of a spin-down effect in BHNSs, our
work also offers numerous opportunities for further studies. First, while
the fit in Fig.~\ref{fig:fitchi} is tuned to the DD2 EOS, its functional
validity will be maintained also for different EOSs, and it remains an
open question whether the spin-down coefficients exhibit quasi-universal
behaviours. Second, given the strong nonlinearity in TDEs and the fact
that the remnant BH spin is maximised for $\chiNS=0$, it is possible that
NSs with anti-aligned spins produce even larger final spins. Finally,
while simple post-Newtonian estimates indicate that the spin-orbit
coupling terms are maximised when the NS spin is aligned, exploring the
BH spin evolution for generic misaligned configurations could provide
novel insights into the nonlinear TDE dynamics and how it differs from
equivalent BBH dynamics. We will explore some of these scenarios in
future work.

\smallskip\noindent\emph{Acknowledgments}---It is a pleasure to thank
P. Singh for comments on the manuscript, and M. Locchi for fruitful
discussions.
Support comes from the ERC Advanced Grant ``JETSET: Launching,
propagation and emission of relativistic jets from binary mergers and
across mass scales'' (Grant No. 884631) and the DFG through the CRC-TR
211 ``Strong-interaction matter under extreme conditions'' (project
number 315477589) TRR 211. F.~C. acknowledges support from the Horizon
Europe staff exchange (SE) programme HORIZON-MSCA2021-SE-01 Grant No.
NewFunFiCO-101086251. L.~R. acknowledges the Walter Greiner Gesellschaft
zur F\"orderung der physikalischen Grundlagenforschung e.V. through the
Carl W. Fueck Laureatus Chair. The simulation were performed on the ITP
Supercomputing Cluster Calea and on the GCS Supercomputer SuperMUC-NG at
the Leibniz Supercomputing Centre (LRZ) in Garching, Germany, under the
project pn67xi.

\bibliography{BHNS_spin.bbl}

\clearpage
\mbox{}
\section{End Matter}
\begin{figure}[h!]
  \centering
  \includegraphics[width=1\columnwidth]{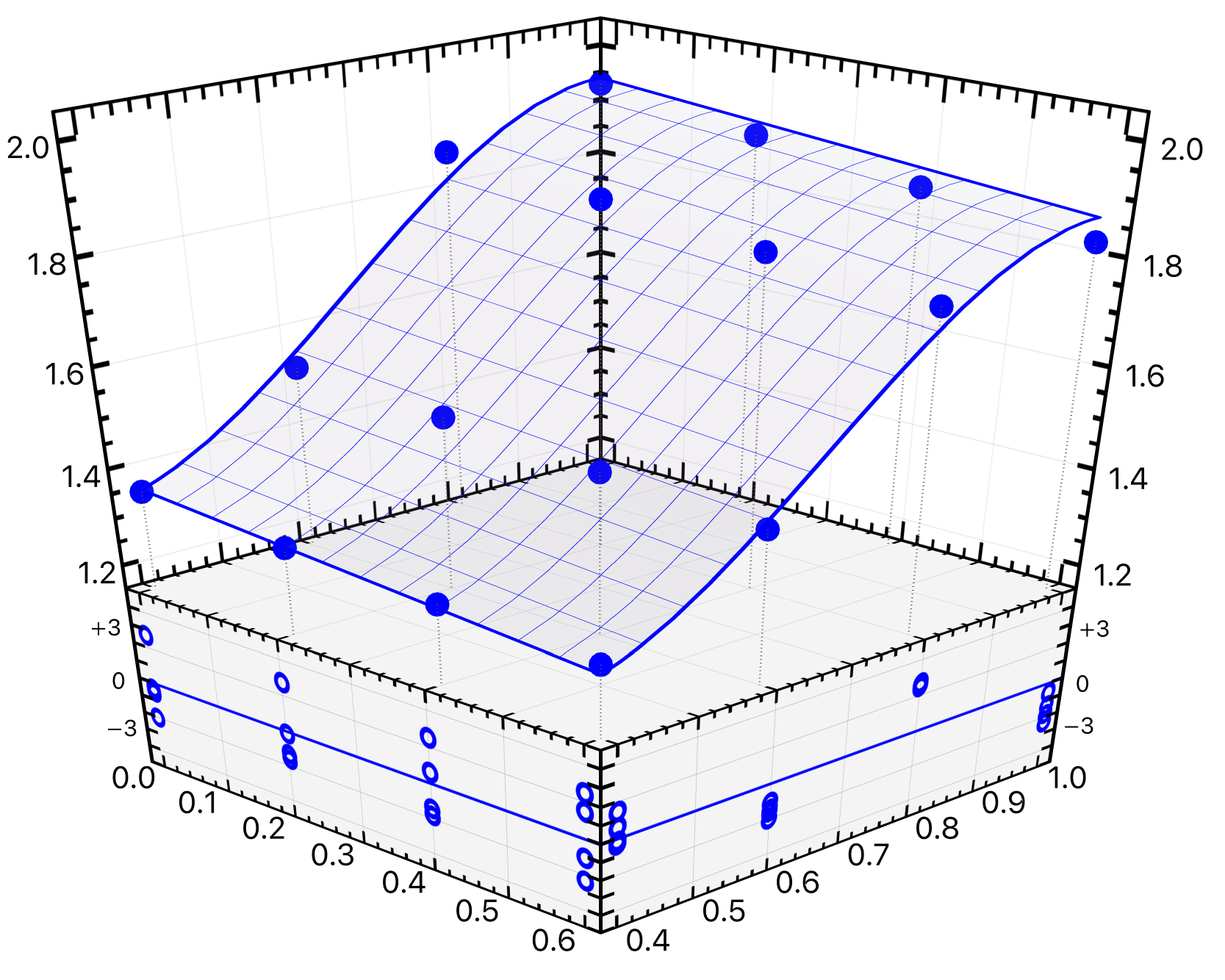}
  \begin{picture}(0, 0)
    \put(-100, 165){\rotatebox{7}{\large{$\frac{\MBH}{\MBHi}$}}}
    \put(-65, 25){\rotatebox{-25}{\large{$\chiNS$}}}
    \put(45, 20){\rotatebox{45}{\large{$q$}}}
    \put(100, 60){\scriptsize\rotatebox{86}{$\delta\tiny{[\%]}$}}
    \put(-110,60){\scriptsize\rotatebox{94}{$\delta\tiny{[\%]}$}}
  \end{picture}
  \caption{The same as Fig.~\ref{fig:fitchi} but for the BH mass.}
  \label{fig:fitM}
\end{figure}

\begin{figure}[h!]
  \centering
  \includegraphics[width=1.0\columnwidth]{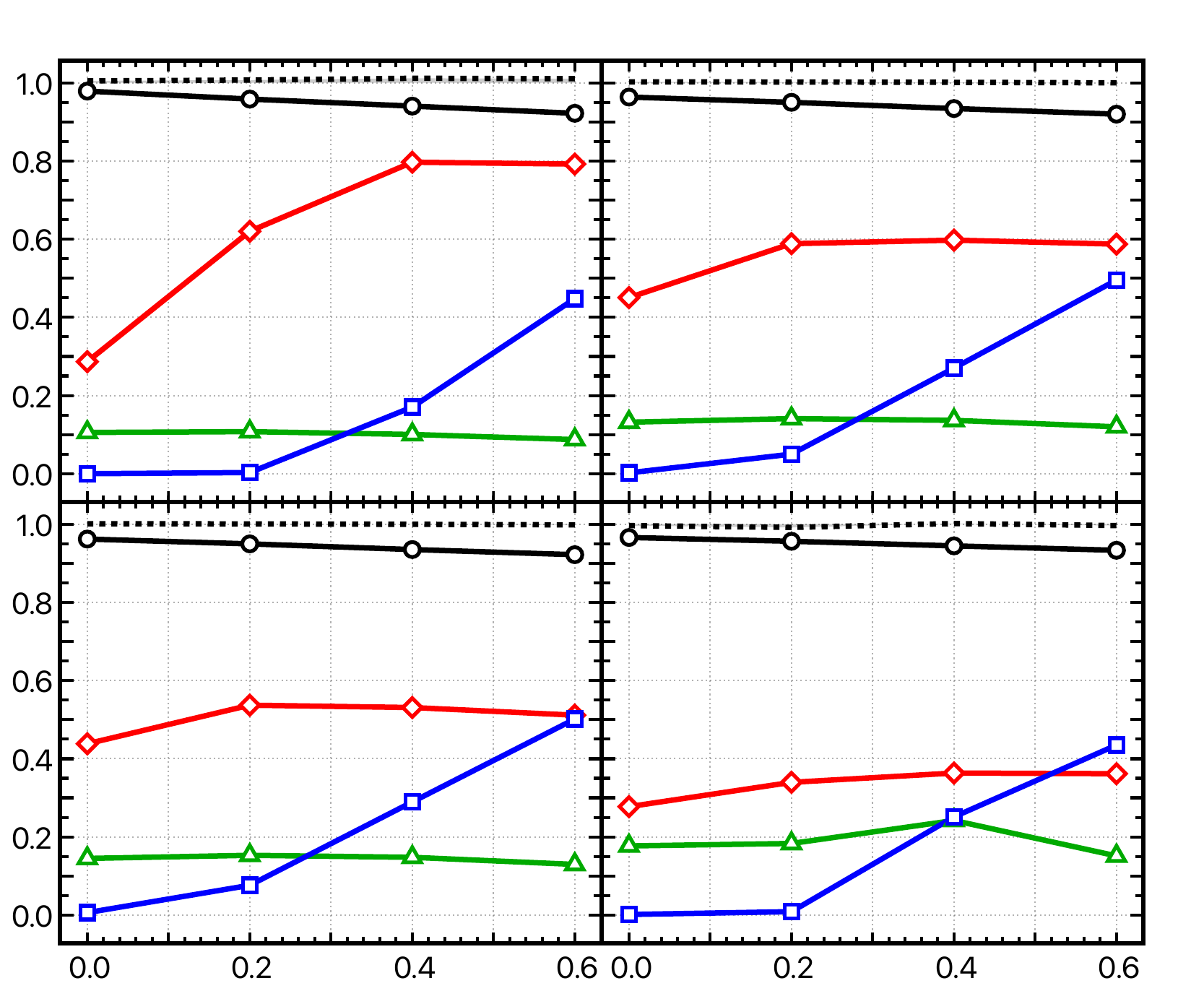}
  \begin{picture}(0, 0)
    \put(-100, 187){\footnotesize {$\tilde{M}_{_{\rm BH}}$}}
    \put(-100, 133){\footnotesize \textcolor{Green}{$15\!\times\!\tilde{E}_{_{\rm GW}}$}}
    \put(-51, 171){\footnotesize
      \textcolor{red}{$15\!\times\!\tilde{M}_{b, {\rm disk}}$}}
    \put(-51, 152){\footnotesize \textcolor{blue}{$15\!\times\!\tilde{M}_{b,{\rm ej}}$}}
    \put(-60, 10){$\chiNS$}
    \put(50, 10){$\chiNS$}
    \put(-36, 117){ {$q=1.0$}}
    \put(75, 117){ {$q=0.8$}}
    \put(-36, 27){ {$q=0.6$}}
    \put(75, 27){ {$q=0.4$}}
  \end{picture}
  \caption{The same as Fig.~\ref{fig:balance} but for the BH mass, with
    $\tilde{M}_{n}:= M_{n}/E_{0}$}
  \label{fig:balanceM}
\end{figure}

In analogy to Fig.~\ref{fig:fitchi}, we report in Fig.~\ref{fig:fitM},
the relative growth of the BH mass at $10\,{\rm ms}$ after merger,
$\MBH/\MBHi$, is plotted as a function of $q$ and $\chiNS$, together with
a polynomial fit \hbox{$\MBH/\MBHi\!=\![k_0\!+\!k_1 q \!+\!k_2 q^2
    \!-\!{(k_1\!+\!2k_2)}q^3/3 \!+\! h_1 \chiNS]$}, where the cubic term
is fixed by demanding the maximum to be at $q=1$ for all $\chiNS$, and
$k_0=1.94\pm 0.28$, $k_1=-4.1\pm 1.2$, $k_2=8.2\pm 1.6$, $h_1=-0.13\pm
0.04$. The fit in this case has relative differences that are $\lesssim
3\%$.
Similarly, in analogy to Fig.~\ref{fig:balance}, we report in
Fig.~\ref{fig:balanceM} the total energy balance, $E_{0}=M_{_{\rm
    BH}}+E_{_{\rm GW}}+M_{b, \rm disk}+M_{b, {\rm ej}}$, with $E_0$ the
initial ADM mass of the system provided by \texttt{FUKA}, $M_{_{\rm
    BH}}$ the mass of the BH remnant at $10\,{\rm ms}$ after merger, and
$E_{_{\rm GW}}$ the cumulative energy radiated in GWs. The energy of the
disk and of the ejecta (magnified by a factor $\times15$ for visibility)
are approximated by their baryon-mass, $M_{b, \rm disk}$ and $M_{b, {\rm
    ej}}$, thus neglecting pressure contribution and self-gravity; clearly
$90\%$ of the total energy is accreted by the BH.
%
\subsection*{Long-term properties of the black-hole remnant}
For BH remnants of compact binary mergers, the spin parameter will
continue evolving primarily via accretion and over viscous
timescales. While the attainment of a MAD state and efficient magnetic
torque braking would spin down the BH in a variety of accretion
scenarios~\cite{Gammie:2003qi, Tchekhovskoy2012, Nathanail_2020,
  Nathanail2021b, Narayan2022, Jacquemin-Ide2023a, Lowell2023,
  Nathanail2025, Issa2026}, there is growing numerical evidence that
disks from compact binaries are unlikely to go MAD within a timescale
that overlaps with peak jet luminosities~\cite{Hayashi2022, Hayashi2023,
  Gottlieb2023a, Hayashi2024}, and indeed a global MAD state might be
achieved as late as $t\gtrsim 10~\rm{s}$~\cite{Izquierdo2025}.
Therefore, we neglect its effects in the following.
In order to establish a quantitative relation between the properties of
the BH remnant self-consistently determined via simulations at $10~{\rm
  ms}$ and its asymptotic properties we define the asymptotic (in time)
BH spin as $\chi_{_{\rm BH}}^{\infty}:= \lim_{t\to\infty}\, \chi_{_{\rm
    BH}} = (J_{_{\rm BH}}+ (1 - \zeta) {J}_{\rm disk})/[{M}_{_{\rm BH}} +
  (1-\xi) {M}_{\rm disk}]^2$, where $\xi$ and $\zeta$ are
phenomenological parameters, introduced to represent the fractions of the
disk mass and angular momentum that are lost at infinity as a consequence
of neutrino and magnetically driven winds.
While determining $\xi$ and $\zeta$ by evolving the system fully
dynamical relativistic simulations for timescales of several seconds is
prohibitively expensive, we consider a robust but simplified model
constructed by assuming a quasi-stationary evolution for the accreting BH
for $t>10\,{\rm ms}$, where the BH mass and angular-momentum changes are
\begin{align}
  \label{eq:evolution_model}
  \frac{d\MBH}{dt}&=\dot{\mathcal{M}}(t)\,, \qquad \frac{dJ_{_{\rm
        BH}}}{dt}=\dot{\mathcal{M}}(t)\ell_{_{\rm ISCO}}(t) \,.
\end{align}
The initial conditions at time $t_0=10\,{\rm ms}$ are provided by $\MBH$
and $J_{_{\rm BH}}$ as measured from simulations (see
Table~\ref{table:1}), and $\ell$ is the specific angular momentum in a
Kerr spacetime $\ell=\pm \MBH (r^2\mp 2\chiBH\sqrt{r}+\chi_{_{\rm
    BH}}^2)/(\sqrt{r}\left(r^2-3r \pm 2\chiBH \sqrt{r}\right)^{1/2})$,
evaluated at the ISCO radius (see, \eg \cite{Rezzolla_book:2013}). The
system of Eqs.~\eqref{eq:evolution_model} requires a prescription for the
mass accretion-rate $\dot{\mathcal{M}}$, which we model as
$\dot{\mathcal{M}}(t) := ({{M}_{{\rm disk}}(t_0)}/{{M}_{_{\rm
      G,H}}(t_0)})\dot{M}_{{\rm ref}}(t)$, where $\dot{M}_{{\rm ref}}(t)$
is extracted from seconds-long simulations in the Cowling approximation
and modelling neutrino cooling from Ref.~\cite{Gottlieb2023} (labelled
with ${\rm G}$) and Ref.~\cite{Hayashi2023} (labelled with ${\rm H}$),
and where the initial disk mass is appropriately rescaled. Upon
integrating Eqs.~\eqref{eq:evolution_model}, we can determine the
fractions $\xi$ and $\zeta$ leading to $\partial \chiBH/\partial \chiNS
<0$, finding $\xi_{_{\rm G}}\approx0.34$ and $\zeta_{_{\rm G}}\approx0.5$
for the model based on \cite{Gottlieb2023}, whereas $\xi_{_{\rm
    H}}\approx0.6$ and $\zeta_{_{\rm H}}\approx0.7$
for~\cite{Hayashi2023}. In both cases, these values present small
fluctuations over the progenitor's mass-ratio and NS spin.

\begin{figure}[h!]
  \centering
  \includegraphics[width=0.66\columnwidth]{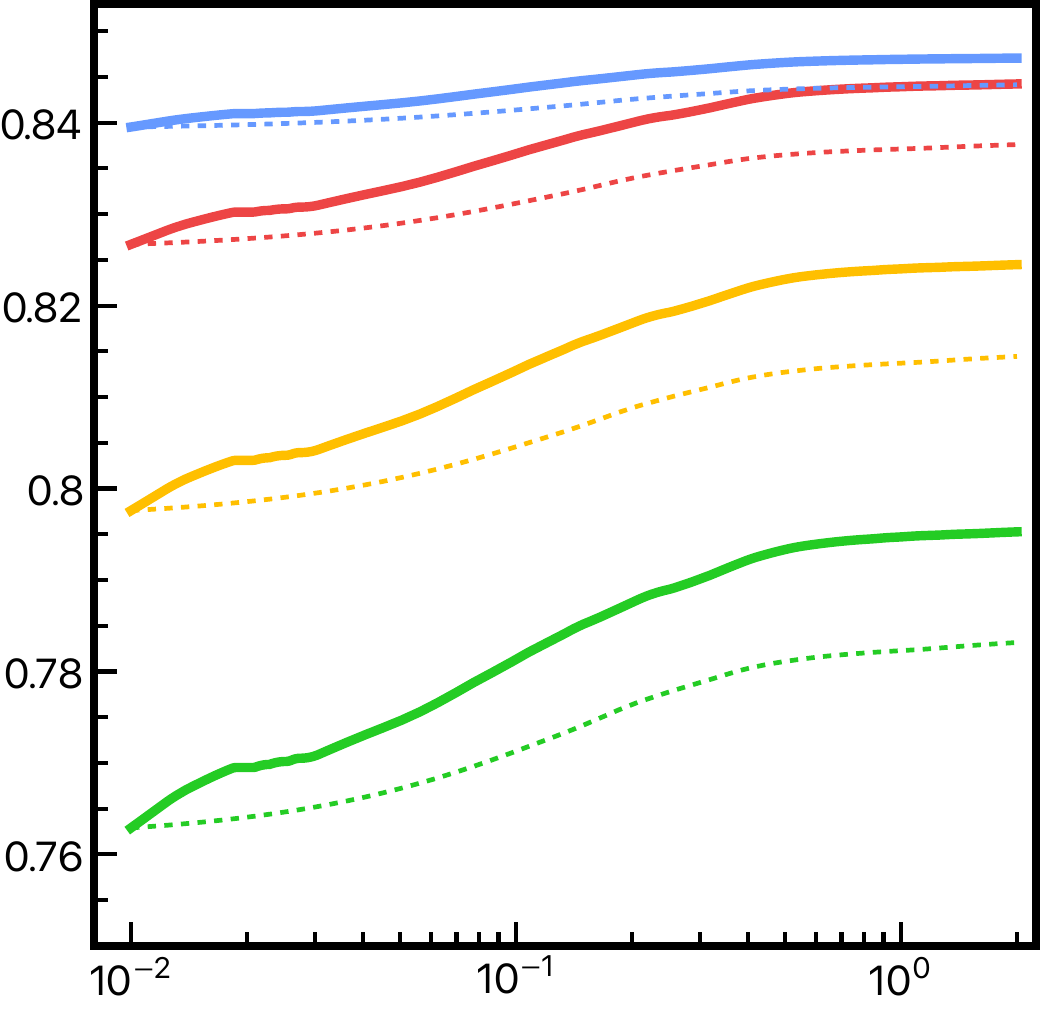 }
  \begin{picture}(0, 0)
    \put(-180, 75){\rotatebox{90}{$\chiBH$}}
    \put(-100, -10){$t-t_{\rm merg}~[{\rm s}]$}
    \put(-60, 40){\color{mathGreen}{$\chiNS=0.6$}}
    \put(-60, 90){\color{mathYellow}{$\chiNS=0.4$}}
    \put(-60, 125){\color{mathRed}{$\chiNS=0.2$}}
    \put(-140, 145){\color{mathBlue}{$\chiNS=0.0$}}
  \end{picture}
  \vspace{3mm}
  \caption{Long-term evolution for the BH remnant spin in the $q=1$ case,
    on viscous-accretion timescales. The solid and dashed line use the
    accretion rates in~\cite{Gottlieb2023} (${\rm G}$),
    and~\cite{Hayashi2023} (${\rm H}$), respectively. Similar curves have
    been also obtained for $q=0.8,\,0.6,\,0.4$.}
  \label{fig:Long_Time}
\end{figure}
In Fig.~\ref{fig:Long_Time} we report the long-term evolution of the BH
spin parameter in the case $q=1$, showing that it preserves the hierarchy
of spin parameters $\partial \chiBH/\partial \chiNS <0$ that we observed
at $10~{\rm ms}$, supporting the non-additivity of spins in BHNS binaries
also on secular timescales. Stated differently, even when enabling disk
accretion over long timescales, it continues to be true that the BH
remnants for BHNS with nonrotating NSs will have a spin larger than those
with rapidly rotating NSs. Similar behaviors have been obtained also for
$q\neq 1$.
\subsection*{Spin corrections in the remnant baryon mass}
\begin{figure}
  \centering
  \includegraphics[width=0.66\columnwidth]{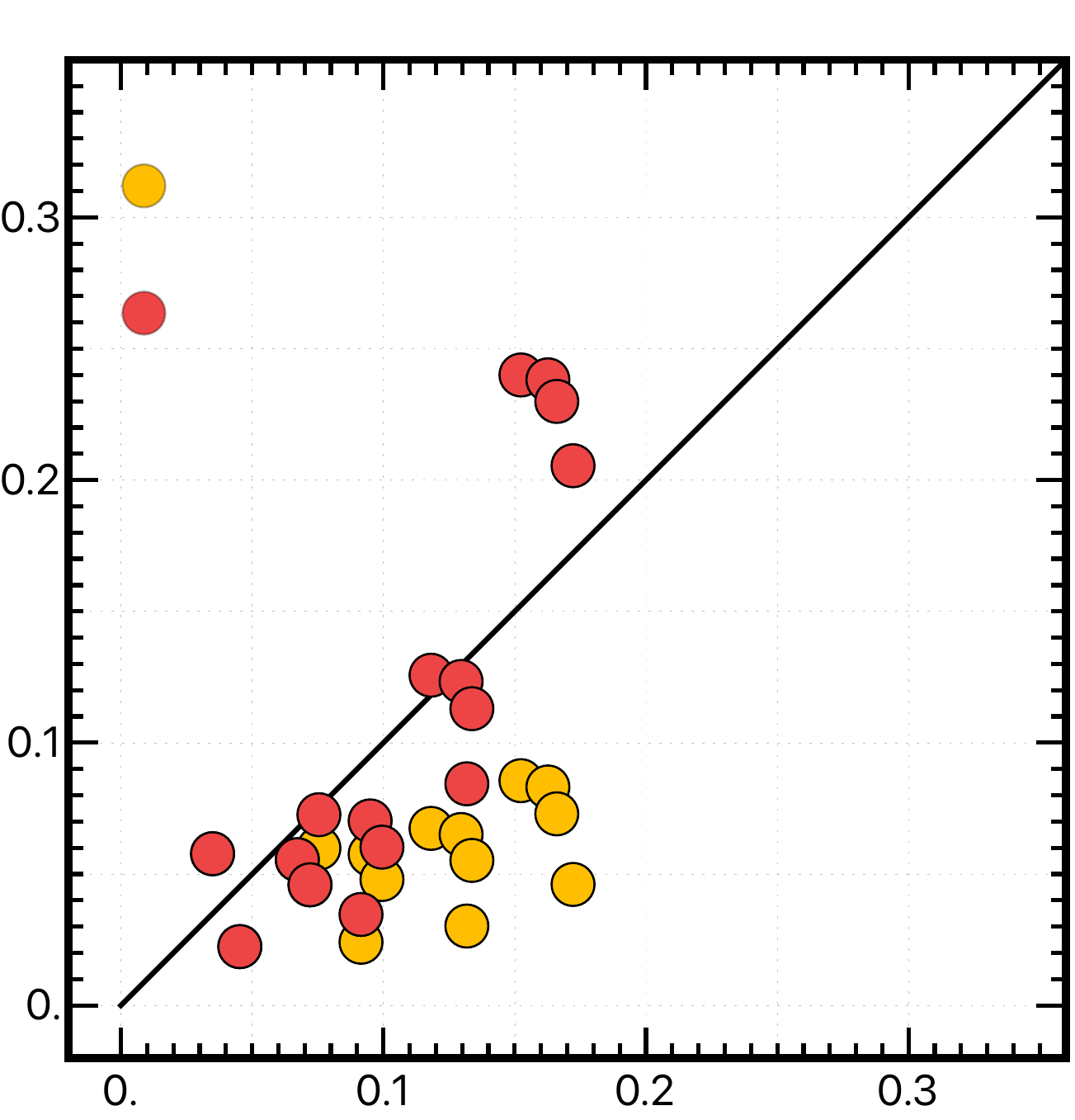}
  \begin{picture}(0, 0)
    \put(-180, 65){\rotatebox{90}{$M_{b, {\rm rem}}/\MBNS$}}
    \put(-115, -10){$M_{b, {\rm rem~(sim)}}/\MBNS$}
    \put(-138, 138){\,${\rm Foucart+~(2018)}$, Eq.~\eqref{eq:FF18}}
    \put(-138, 118){\,${\rm Spin}$-${\rm correction}$, Eq.~\eqref{eq:remnantbMassfit}}
  \end{picture}
  \vspace{3mm}
  \caption{Comparison between the total remnant baryon-mass from our
    simulations from previous predictions Eq.~\eqref{eq:FF18} (yellow
    circles) and from our spin-corrected improved expression
    Eq.~\eqref{eq:remnantbMassfit} (red circles).}
  \label{fig:FF18_comparison}
\end{figure}

The baryon-mass and the angular momentum of the disk (ejecta) are
computed via \hbox{$M_{b, {\rm disk~(ej)}} := -\int_{\mathcal{V}_{{\rm
        disk~(ej)}}} \rho n_{\mu}u^\mu \sqrt{\gamma}d^{3}x$} and
\hbox{$J_{b, {\rm disk~(ej)}} := \int_{\mathcal{V}_{{\rm disk~(ej)}}}
  n_{\mu} (\partial_{\phi})^{\nu} T^{\mu}_{~\nu}\sqrt{\gamma}d^{3}x$},
where $\rho$ and $u^\mu$ are the the fluid density and $4$-velocity,
respectively, whereas $\mathcal{V}_{\rm disk}$ and $\mathcal{V}_{\rm ej}$
respectively label the volumes comprised within a sphere of
$r=900~M_\odot$ and constrained by $hu_t>-1$ and $h u_t<-1$.
The remnant baryon-mass at $10\,{\rm ms}$ after merger, \hbox{$M_{b,{\rm
      rem}} := M_{b,{\rm disk}}+M_{b, {\rm ej}}$} constitutes a
fundamental quantity for multimessenger astrophysics, directly related to
electromagnetic counterparts in binary mergers, and theoretical
predictions from numerical-relativity are often expressed in terms of the
fitting function~\cite{Foucart2012, Foucart2018b}
\begin{equation}
  \frac{M_{b, {\rm rem}}}{\MBNS}=\left[{\rm
      max}\left(\alpha\frac{1-2\mathcal{C}}{\eta^{1/3}} - \beta
    \frac{\mathcal{C}}{\eta}\frac{r_{_{\rm
          ISCO}}}{\MBH}+\gamma,~0\right)\right]^\delta \,,
  \label{eq:FF18}
\end{equation}
where $\eta:=q/(1+q)^2$ is the symmetric mass-ratio, $\alpha=0.406$,
$\beta=0.139$, $\gamma=0.255$ and $\delta=1.761$. We note that the fit
\eqref{eq:FF18} was built using nonrotating NSs only and that an obvious
ambiguity emerges when adopting \eqref{eq:FF18} for rotating NS, since
the compactness can be defined in terms of the areal radius,
$\mathcal{C}$, or of the equatorial radius, $\mathcal{C}_{\rm eq}$ (see
caption in Table~\ref{table:1}).
Clearly, Eq.~\eqref{eq:FF18} needs to be corrected to account for the
different (smaller) compactness that accompanies rapidly rotating
stars. To this scope, we can leverage on the quasi-universal relation
between the polar and equatorial proper radii~\cite{Musolino2023b}
${R_{\rm pol}}/{R_{\rm eq}} := 1 - \varepsilon ({\chiNS}/{\chi_{\rm
    Kep}})^2$, where $\chi_{\rm Kep}$ is the maximum Keplerian
dimensionless spin for a given EOS, which can be approximated as
$\chi_{\rm Kep} \simeq b_4 \sqrt{{R_{_{\rm TOV}}}/{M_{_{\rm TOV}}}}
\simeq 0.69$~\cite{Breu2016}, with $b_4 \simeq 0.554$, and {$M_{_{\rm
      TOV}}=2.421\,M_{\odot}, R_{_{\rm TOV}}=8.066\,{\rm km}$} for the
DD2 EOS~\footnote{We should remark that the coefficient $\varepsilon$ was
indicated as ``$b_2 \approx 0.40$'' in~\cite{Musolino2023b}, but this
symbol has been used already here when expressing the fitting expression
for $\chiBH$.}. We can now improve the baryonic-mass prediction by
replacing $\mathcal{C}$ in Eq.~\eqref{eq:FF18} by the equatorial
compactness $\mathcal{C}_{\rm eq}$ defined as $\mathcal{C}_{\rm eq} :=
\mathcal{C} [1 - \varepsilon ({\chiNS}/{\chiKep})^2]$. In this way, we
can effectively extend the regime of validity of previous fits to
low-mass BHNS~\cite{Foucart2012, Foucart2018b} expressed by
Eq.~\eqref{eq:FF18} to include also spin-induced corrections:
\begin{align}
  \nonumber
  & \hskip -1.0cm \frac{M_{b, {\rm rem}}}{\MBNS}=\left[{\rm
      max}\left(\alpha\frac{1-2\mathcal{C}(1 -
      \varepsilon \chiNS^2/\chiKep^2)}{\eta^{1/3}}\right.\right.
    \\
    &\quad\left.\left.-\beta \frac{\mathcal{C}
      (1-\varepsilon\chiNS^2/\chiKep^2)}{\eta}\frac{r_{_{\rm
          ISCO}}}{\MBH}+\gamma,~0\right)\right]^\delta \,.
  \label{eq:remnantbMassfit}
\end{align}
Figure~\ref{fig:FF18_comparison} compares the remnant baryon mass, $M_{b,
  {\rm rem~(sim)}} := M_{b, {\rm disk}} + M_{b,{\rm ej}}$ from our
simulations against the previous fitting formula, Eq.~\eqref{eq:FF18}
(yellow circles), and the new spin-extended prescription,
Eq.~\eqref{eq:remnantbMassfit} (red circles), showing that the latter
compensates for a systematic underestimation of the numerical data (if
left free, the best-fit value is $\varepsilon_{\rm best~fit} \simeq
0.63$, which is interestingly close to what predicted from
quasi-universal relations, $\varepsilon \simeq
0.41$~\cite{Musolino2023b}).
It should be noted that expression~\eqref{eq:remnantbMassfit} represents
a significant improvement over the previous fit~\eqref{eq:FF18} without
additional fitting coefficients, since $\varepsilon$ is not a free
parameter but is derived from universal
relations~\cite{Musolino2023b}. It remains interesting to improve the new
fit against the complete dataset of BHNSs simulated so far, either in the
well-known regime of $q \leq 0.2, \chiNS \simeq 0$, but also, and
especially, in the regime of $q \lesssim 1.0, \chiNS \simeq
\chiKep$~\cite{Ruiz2020b}.

\clearpage
\mbox{}
\section{Supplemental Material}

\setcounter{figure}{0}
\setcounter{equation}{0}

\subsection*{Numerical Setup}
We here provide some details on the numerical aspects of the simulations
and on the quantities measured.

The initial data for quasi-circular BHNS sequences are produced with the
publicly available spectral solver
\texttt{FUKA}~\cite{Grandclement:2009ju, Papenfort2021b, Topolski2024b},
at a sufficiently high, but otherwise computationally manageable,
resolution of $N=(13, 13, 12)$ collocation points in all directions
across all spectral domains. For each sequence, we fix the NS
gravitational mass to a reference value $M_{_{\rm NS}}=1.35\, M_{\odot}$,
but we vary the BH mass to obtain binaries with mass ratio $q=[0.4, 0.6,
  0.8, 1.0]$, and progressively larger NS spin aligned with the orbital
angular momentum, $\chi_{_{\rm NS}}=[0, 0.2, 0.4, 0.6]$. Note that
$\chiNS=0.2$ and $\chiNS=0.6$ correspond to spin frequencies $\approx
302\, \rm{Hz}$ and $\approx 760\, \rm{Hz}$, respectively, which are close
to the inferred upper limits for some of the fastest spinning
pulsars~\cite{Lynch2012, Hessels2006}. The number of orbits before merger
depends obviously on $q$ and $\chiNS$, and we found that an initial
separation $d_{0}=12 (\MNS+\MBH)$ represents an optimal choice to provide
a sufficient number of cycles and a residual eccentricity that is
comparable or below $10^{-3}$. It is worth noting that BHNS binaries with
light BHs (in some cases even with sub-solar masses~\cite{Foucart2019,
  Markin2023}) are receiving widespread attention driven by recent
detections~\cite{Han2020, Qin2024, Matur2026} such as
GW230529~\cite{LIGOScientific2024} and GW190425~\cite{Abbott2020}, and
because they act as BNS mimickers~\cite{Hinderer2018, Foucart2018b,
  Foucart2019, Hayashi2020, Most2020e, Markin2026}.

All binaries are evolved over a domain of extent $\pm 1500\, \rm{km}$ and
$z$-reflection symmetry covered by $9$ levels of refinement, where the
finest resolution has a spacing $dx=0.144\, M_{\odot} \simeq 213\,
\rm{m}$ (medium resolution, MR). Two additional resolutions $dx=0.12\,
M_{\odot} \simeq 177\, \rm{m}$ (high, HR) and $dx=0.1\, M_{\odot} \simeq
148\, \rm{m}$ (very high, VHR) have been employed to evolve the most
critical configuration with $q=1,\, \chiNS=0.0$, while a simulations at
HR has been performed to measure the accuracy in the binary $q=1,\,
\chiNS=0.2$ (see Table~\ref{table:1}).
\begin{table}
    \centering
    \setlength{\tabcolsep}{1.5pt}
    \renewcommand{\arraystretch}{0.5} 
    \small
    \begin{tabular}{|l|c|c|c|c|c|c|c|c|c|}
  \hline
  \rule{0pt}{2ex} {\scriptsize Res.} & $q$ & $\chiNS$ & $E_0$ & $J_{0}$ & $\MBHi$ & $M_{\rm disk}$ & $M_{\rm ej}$ & $\MBH$ & $\chiBHf$ \\
  \hline
 & & & {\scriptsize $[M_{\odot}]$}
  & {\scriptsize $[M_{\odot}^{2}]$}
  & {\scriptsize $[M_{\odot}]$}
  & {\scriptsize $[M_{\odot}/10^2]$}
  & {\scriptsize $[M_{\odot}/10^2]$}
  & {\scriptsize $[M_{\odot}]$}
  & \\
  \hline
  {\scriptsize ${\rm VHR}$} & $1.0$ & $0.0$ & $2.67$ & $7.46$ & $1.35$ & $5.1$  & $<0.1$ & $2.62$ & $0.84$ \\
  {\scriptsize ${\rm HR} $} & $1.0$ & $0.2$ & $2.67$ & $7.79$ & $1.35$ & $11.1$ & $<0.1$ & $2.56$ & $0.83$ \\
  {\scriptsize ${\rm MR} $} & $1.0$ & $0.4$ & $2.67$ & $8.13$ & $1.35$ & $14.2$ & $~3.0$ & $2.52$ & $0.80$ \\
  {\scriptsize ${\rm MR} $} & $1.0$ & $0.6$ & $2.68$ & $8.47$ & $1.35$ & $14.1$ & $~8.0$ & $2.47$ & $0.76$ \\
  \hline
  {\scriptsize ${\rm MR} $} & $0.8$ & $0.0$ & $3.27$ & $10.78$ & $1.69$ & $~9.8$ & $~0.1$ & $3.15$ & $0.75$ \\
  {\scriptsize ${\rm MR} $} & $0.8$ & $0.2$ & $3.27$ & $11.11$ & $1.69$ & $12.8$ & $~1.1$ & $3.11$ & $0.74$ \\
  {\scriptsize ${\rm MR} $} & $0.8$ & $0.4$ & $3.27$ & $11.44$ & $1.69$ & $13.0$ & $~5.9$ & $3.05$ & $0.71$ \\
  {\scriptsize ${\rm MR} $} & $0.8$ & $0.6$ & $3.27$ & $11.77$ & $1.69$ & $12.8$ & $10.8$ & $3.01$ & $0.67$ \\
  \hline
  {\scriptsize ${\rm MR} $} & $0.6$ & $0.0$ & $3.57$ & $12.43$ & $2.25$ & $10.4$ & $~0.2$ & $3.43$ & $0.71$ \\
  {\scriptsize ${\rm MR} $} & $0.6$ & $0.2$ & $3.57$ & $12.77$ & $2.25$ & $12.8$ & $~1.8$ & $3.49$ & $0.70$ \\
  {\scriptsize ${\rm MR} $} & $0.6$ & $0.4$ & $3.57$ & $13.09$ & $2.25$ & $12.6$ & $~6.9$ & $3.34$ & $0.67$ \\
  {\scriptsize ${\rm MR} $} & $0.6$ & $0.6$ & $3.57$ & $13.43$ & $2.25$ & $12.2$ & $11.9$ & $3.29$ & $0.63$ \\
  \hline
  {\scriptsize ${\rm MR} $} & $0.4$ & $0.0$ & $4.73$ & $18.67$ & $3.38$ & $~8.7$ & $~0.5$ & $4.53$ & $0.60$ \\
  {\scriptsize ${\rm MR} $} & $0.4$ & $0.2$ & $4.69$ & $18.99$ & $3.38$ & $10.6$ & $~2.8$ & $4.49$ & $0.58$ \\
  {\scriptsize ${\rm MR} $} & $0.4$ & $0.4$ & $4.69$ & $19.31$ & $3.38$ & $11.4$ & $~7.9$ & $4.43$ & $0.56$ \\
  {\scriptsize ${\rm MR} $} & $0.4$ & $0.6$ & $4.69$ & $19.65$ & $3.38$ & $11.4$ & $13.6$ & $4.38$ & $0.53$ \\
  \hline
\end{tabular}
    \caption{The initial data are produced with \texttt{FUKA}, holding
      $\MNS=1.35\, M_{\odot}$ fixed. $J_{0}$ is the initial ADM angular
      momentum, corresponding to the separation $d_{0}=12 (\MNS+\MBHi
      )$. The baryon-mass and the stellar compactness vary with $\chiNS$,
      \ie $M_{b,{_{\rm NS}}}/ M_{\odot}\simeq [ 1.469 , 1.466, 1.459,
        1.452]$ and $\mathcal{C}_{\rm eq}\simeq[0.152, 0.149, 0.142,
        0.128]$, the latter computed via the equatorial/circumferential
      NS radius, whereas $\mathcal{C}\simeq[0.152, 0.151, 0.148, 0.141]$,
      is computed via the areal NS radius.}
    \label{table:1}
\end{table}

The energy and angular momentum cumulatively radiated away in GWs are
computed in terms of the $(\ell,m)$-modes of the strain, $h_{\ell
  m}$~(see, \eg \cite{Bishop2016} for explicit expressions) with the
detector placed at $r=900~M_\odot$ and where we consider
$\ell=2,3,4$. The ADM mass and angular momentum computed by \texttt{FUKA}
are evaluated as surface integrals at spatial infinity (see
\cite{Papenfort2021b} for explicit expressions). The BH properties are
tracked via quasi-local measurements based on the isolated-horizon
framework~\cite{Ashtekar01a, Dreyer02a}, $J_{_{\rm
    BH}}:=({1}/{8\pi})\int_{S_{\rm AH}} (\varphi^\mu R^\nu K_{\mu\nu})
dV^2$, and $M_{_{\rm BH}}:={\sqrt{r_{_{\rm BH}}^4+4J_{_{\rm
        BH}}^2}}/{r_{_{\rm BH}}}$ where $dV$ is the surface element on
the apparent-horizon (AH), $R^\mu$ is the unit outward-pointing spacelike
normal to the AH worldtube, $\varphi^\mu$ the tangent vector, and
$r_{_{\rm BH}}$ is the AH radius. Using this infrastructure, we estimate
the fitting coefficients for the final BH spin and the final BH mass for
the DD2 EOS.

\subsection*{Gravitational-wave signals}
%
\begin{figure*}[t!]
  \centering \includegraphics[width=0.85\textwidth]{
    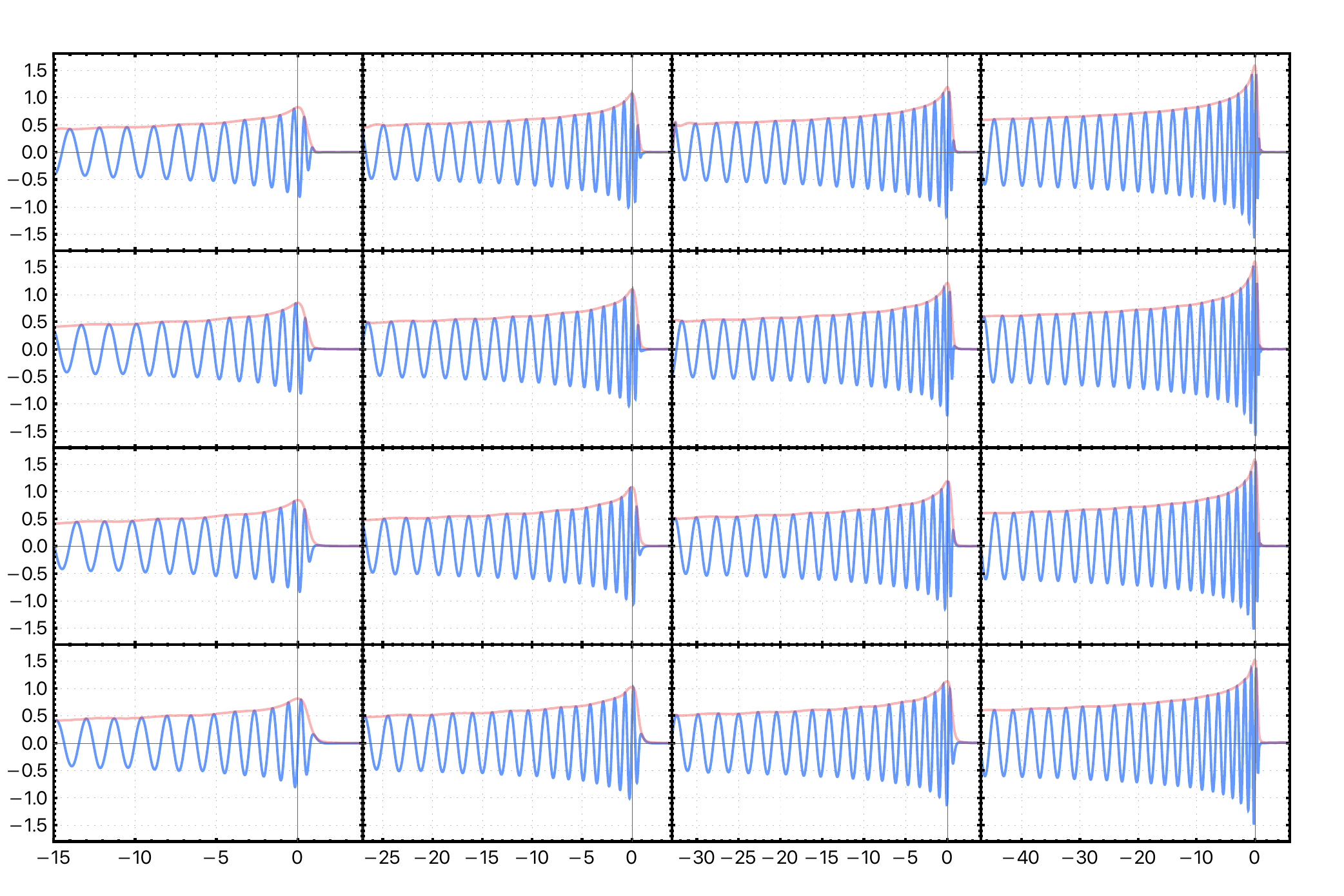 }
  \begin{picture}(0, 0)
    \put(-380, 280){$q=1.0$}
    \put(-280, 280){$q=0.8$}
    \put(-190, 280){$q=0.6$}
    \put(-85, 280){$q=0.4$}
    \put(-405, 262){$\chiNS=0.0$}
    \put(-405, 200){$\chiNS=0.2$}
    \put(-405, 135){$\chiNS=0.4$}
    \put(-405, 72){$\chiNS=0.6$}
    \put(-90, -3){$t-t_{\rm merg}~[\rm ms]$}
    \put(-200, -3){$t-t_{\rm merg}~[\rm ms]$}
    \put(-290, -3){$t-t_{\rm merg}~[\rm ms]$}
    \put(-390, -3){$t-t_{\rm merg}~[\rm ms]$}
    \put(-450, 110){\rotatebox{90}{$10^{24}\times h_{22}^{+}~[40\,{\rm Mpc}]$}}
  \end{picture}
  \vspace{3mm}
  \caption{GW waveforms measured at $40\, Mpc$ from the source, for
    the dominant $(\ell, m)=(2, 2)$ modes measured at $r=900
    M_\odot$. Blue is used for the $h_+$ polarization (real part of
    the strain), whereas the red shaded curve represents its amplitude
    envelope. Each column collects binary configurations at fixed
    mass-ratio $q=[1.0,0.8,0.6,0.4]$, whereas each row is
    characterised by a NS spinning at $\chiNS=[0.0,0.2,0.4,0.6]$. All
    the waveforms reported here correspond to MR numerical
    simulations.}
  \label{fig:GW_waveforms}
\end{figure*}
We provide here details over the GW signals extracted from our
simulation campaign.
We read the Weyl scalar $\psi_4$ on a spherical-surface placed at a 
distance $900\,M_\odot\approx 1300 \, {\rm Km}$ from the center of the
grid, and extract the GW strain via spin-weighted spherical harmonics
decomposition
\begin{equation}
  h^++ih^\times:=\sum_{\ell=2}^\infty\sum_{m=|\ell|} h_{\ell m}\,_{_{-2}}Y_{\ell m}\,,
\end{equation}
with the mode-expansion that in our simulations truncates at the
hexadecapole order, $\ell=4$.
From the GW strain the energy and angular momentum cumulatively
radiated away in GWs can be computed as~\cite{Bishop2016}
\begin{align}
  \label{eq:EGWs}
  E_{_{\rm GW}}(t)&:=-\frac{r^2}{16\pi}\int_{t_0}^t
  \sum_{\ell,m}|\dot{h}_{\ell m}(t')|^2dt'\,, \\
  \label{eq:JGWs}
  J_{_{\rm
      GW}}(t)&:=\frac{r^2}{16\pi}\int_{t_0}^t \sum_{\ell,m}m
  \Im\left[{h}_{\ell m}(t')\dot{h}^*_{\ell m}(t')\right]dt'\,,
\end{align}
where a dot indicates a time derivative. Expressions~\eqref{eq:EGWs}
and~\eqref{eq:JGWs} have been used in order to estimate the GW
contributions in the total energy and angular momentum balance
respectively depicted in Fig.~\ref{fig:balanceM} and
Fig.~\ref{fig:balance}.

We now focus our attention on the $(\ell,m)=(2,2)$ mode, which is the
dominant one for quasi-circular binaries. In Fig.~\ref{fig:GW_waveforms}
we report for each binary configuration in our dataset the real part of
the dominant strain component, $h_{22}^+$ (blue), together with its
amplitude $h_{22}:=\sqrt{(h_{22}^+)^2+(h_{22}^\times)^2}$ (red shaded
curve), with each signal shifted in time by $t_{\rm merg}$.
We remind the reader that our dataset is constructed by fixing the
initial separation at $12 (\MBHi+\MNS)$, so that the distance, and
thus the number of cycles increases with the mass-ratio $q$. Similar
to \cite{Ruiz2020b} we also observe that aligned NS spins cause a
shift in the GW phase and the tendency to increase the number of GW
cycles, consistent with a repulsive spin-orbit coupling.
All the waveforms presented in Fig.~\ref{fig:GW_waveforms} are
characterised by a long inspiral with a large number of cycles
($\approx 10$ for $q=1$ and $\approx 20$ for $q=0.4$). We envision
this makes our waveforms suitable for the construction of hybrid
waveform models to extend the signal in the inspiral phase by
appending post-Newtonian models~\cite{Etienne2012}, and to assess the
detectability of low-mass BHNS in GW detectors, as well as the
distinguishability of the NS spin in these type of systems.

\end{document}